\documentclass[12pt]{article}
\usepackage{amsmath,amssymb,bm}
\usepackage{cite}  
\usepackage{hyperref}
\hypersetup{colorlinks=false}
\title{Scalar particles emission from black holes with topological defects using Hamilton-Jacobi method}
\date{}
\author{\bf Kimet Jusufi\footnote{E-mail: kimet.jusufi@unite.edu.mk}}

\makeatletter
 
  \@addtoreset{equation}{section}
 \makeatother

\begin{document}
\maketitle

\centerline{\it Department of Physics, State University of Tetovo, Ilinden Street nn, 1200, Macedonia}

\vskip 0.5 truecm

\abstract{We study quantum tunneling of charged and uncharged scalar particles from the event horizon of Schwarzschild-de Sitter and Reissner-Nordstr\"{o}m-de Sitter black holes pierced by an infinitely long spinning cosmic string and a global monopole. In order to find the Hawking temperature and the tunneling probability we solve the Klein-Gordon equation by using the Hamilton-Jacobi method and WKB approximation. We show that Hawking temperature is independent of the presence of topological defects in both cases.}

\section{Introduction}
In 1970s, Steven Hawking showed that if we take into account quantum effects near strong gravitational fields leads to thermal radiation from black holes \cite{hawking}. This discovery also suggests a new deep connection between thermodynamics and general relativity. Namely, every black hole has entropy, which is proportional to the surface area of the black hole. Over the years a number of different approaches were introduced in order to calculate the Hawking temperature. Among others, the Wick rotation method \cite{gibbons,gibbons1}, quantum tunneling \cite{perkih}, anomaly method \cite{iso}, and the technique of dimensional reduction \cite{umetsu}. 

The tunneling method treated Hawking radiation as a tunneling process using a semi-classical WKB approximation, where the particle can quantum mechanically tunnel through the horizon and it is observed at infinity as a real particle. The tunneling rate is related to the imaginary part of the action in the classically forbidden region. Generally, there are two methods to obtain the imaginary part of the action. In the first method, the imaginary part of the action is calculated by integrating the radial momentum of the particles \cite{perkih}, in the second method \cite{angheben,srinivasan}, the imaginary part of the action is obtained by solving the relativistic Hamilton-Jacobi equation. Using this method, Hawking radiation has been studied for scalar particles as well as for Dirac particles \cite{ahmed}, in different spacetime configurations.

The aim of this paper is to use Hamilton-Jacobi method and solve the Klein-Gordon equation for scalar charged/uncharged particles from Schwarzschild-de Sitter (SdS) and Reissner-Nordstr\"{o}m-de Sitter (RNdS) black holes in background spacetimes with a spinning cosmic string and a global monopole and extend the results presented here \cite{rahman,rahman1}. Cosmic strings are one dimensional object, that may have been produced by the phase transition in the early universe. A spinning cosmic string is characterized by the rational parameter $a$ and the angular parameter $J$, given by $a =4J$. The spacetime of a cosmic string is locally flat, however, globally is conical due to the deficit angle $\delta\Phi=8\,\pi \,G \,\mu $. Cosmic string can act as a gravitational lens \cite{gott}, it can induce a finite electrostatic self-force on an electric charged particle \cite{linet}, shifts in the energy levels of a hydrogen atom \cite{bezerra}, they were also suggested as an explanation of the anisotropy of the cosmic microwave background radiation. On the other hand, the spacetime geometry of a global monopole of the surface $\theta=\pi/2$, is conical, similar to a cosmic string  with deficit angle $\delta \Omega=8\pi^{2}G\eta^{2}$.

This paper is organised as follows. In Section 2, we briefly review and introduce the line element of SdS black hole with topological defects near the event horizon. In Section 3, we calculate the tunneling rate and Hawking temperature for SdS black hole. In Section 4, we calculate the tunneling rate and Hawking temperature for RNdS black hole. In Section 5, we comment on our results.

\section{SdS black hole with topological defects}
Recall that the simplest Lagrangian density, which describes a global monopole is given by
\begin{equation}
\mathcal{L}=\frac{1}{2}g^{\mu \nu}\partial_{\mu}\phi^{a}\partial_{\nu}\phi_{a}-\frac{1}{4}\lambda (\phi^{a}\phi_{a}-\eta^{2})^{2},
\end{equation}
where $\phi^{a}$ $ (a=1, 2,3) $, is a triplet of scalar fields which transform under the group $O (3) $, whose symmetry is spontaneously broken to $U (1) $, $\lambda$ is the self-interaction term and $\eta$ is the mass term. One can introduce the rotation of an infinitely long cosmic string by simply doing the transformation $\mathrm{d}t\to\mathrm{d}t+a\,\mathrm{d}\phi$ \cite{muniz,vilenkin,vilenkin1}. In this paper, we will consider an idealized cosmic string with a parameter $a$ constant with time, related to the angular parameter $J$, with $a=4J$. Therefore, the line element of the Schwarzschild-de Sitter black hole with positive cosmological factor $\Lambda$ pierced by an infinitely long spinning cosmic string and a global monopole is given by
\begin{eqnarray}\nonumber
\mathrm{d}s^{2}&=&-\left(1-\frac{2M}{r}-\frac{r^{2}}{l^{2}}\right)(\mathrm{d}t+a\,\mathrm{d}\phi)^{2}+\left(1-\frac{2M}{r}-\frac{r^{2}}{l^{2}}\right)^{-1}\mathrm{d}r^{2}
\\
&+& r^{2}p^{2}\left(\mathrm{d}\theta^{2}+b^{2}\sin^{2}\theta\mathrm{d}\phi^{2}\right)
\label{1}
\end{eqnarray}
where $l^{2}=3/\Lambda^{2}$. The first term $p^{2}=1-8\pi\eta^{2}$ encodes the presence of a global monopole, while $b^{2}=(1-4\mu)^{2}$ encodes the presence of a spinning cosmic string characterised by the rational parameter $a$. Solving $r^{3}+2Ml^{2}-rl^{2}=0$, we get the black hole event horizon $r_{H}$ and cosmological horizon $r_{C}$, given by
\begin{equation}
r_{H}=\frac{2M}{3\Xi}\cos\frac{\pi+\psi}{3},\label{3}
\end{equation}
\begin{equation}
r_{C}=\frac{2M}{3\Xi}\cos\frac{\pi-\psi}{3},
\end{equation}
where
\begin{equation}
\psi=\cos^{-1}(3\sqrt{3\Xi}).
\end{equation}

Here $\Xi=M^{2}/l^{2}$, and belongs to the interval $0<\Xi<1/27$. Expanding $r_ {H} $ in terms of $M$ with $\Xi<1/27$, leads to (see, e.g., \cite{rahman})
\begin{equation}
r_{H}=2M\left(1+\frac{4M^{2}}{l^{2}}+\hdots\right),\label{6}
\end{equation}
in the limit $\Xi\to 0$, it follows that $r_{H}\to 2M$. It is more convenient, to write the metric \eqref{1} near the event horizon. For that purpose, one can define $\Delta=r^{2}-2Mr^{2}-r^{4}/l^{2}$, so the line element near the event horizon becomes (\cite{rahman})
\begin{eqnarray}\nonumber
\mathrm{d}s^{2}&=&-\frac{\Delta_{,r}(r_{H})(r-r_{H})}{r^{2}_{H}}(\mathrm{d}t+a\,\mathrm{d}\phi)^{2}+\frac{r^{2}_{H}}{\Delta_{,r}(r_{H})(r-r_{H})}\mathrm{d}r^{2}\\
&+&r_{H}^{2}p^{2}\left(\mathrm{d}\theta^{2}+b^{2}\sin^{2}\theta\mathrm{d}\phi^{2}\right)\label{7}
\end{eqnarray}
where 
\begin{equation}
\Delta_{,r}(r_{H})=\frac{\mathrm{d}\Delta}{\mathrm{d}r}\bigg|_{r=r_{H}}=2\left(r_{H}-M-2\frac{r_{H}^{3}}{l^{2}}\right).
\end{equation}

Since there exists a frame-dragging effect of the coordinate system in the stationary rotating spacetime, we can perform the dragging coordinate transformation $\varphi=\phi-\Omega t$, where
\begin{equation}
\Omega=\frac{a\,\Delta_{,r}(r_{H})(r-r_{H})}{r_{H}^{4}p^{2}b^{2}\sin^{2}\theta-a^{2}\Delta_{,r}(r_{H})(r-r_{H})}.
\end{equation}

In this way the metric \eqref{1} can be written in a more compact form
\begin{equation}
\mathrm{d}s^{2}=-A(r)\mathrm{d}t^{2}+\frac{1}{B(r)}\mathrm{d}r^{2}+C^{2}(r)\mathrm{d}\theta^{2}+D^{2}(r)\mathrm{d}\varphi^{2},\label{10}
\end{equation}
where
\begin{align}
A(r)&=\frac{b^{2}p^{2}r_{H}^{2}\sin^{2}\theta \Delta_{,r}(r_{H})(r-r_{H})}{b^{2}p^{2}r_{H}^{4}\sin^{2}\theta-a^{2}\Delta_{,r}(r_{H})(r-r_{H})}, \\
 B(r)&=\frac{\Delta_{,r}(r_{H})(r-r_{H})}{r^{2}_{H}},    \\
C^{2}(r)&=p^{2}r_{H}^{2},\\
D^{2}(r)&=p^{2}b^{2}r_{H}^{2}\sin^{2}\theta-a^{2}\frac{\Delta_{,r}(r_{H})(r-r_{H})}{r_{H}^{2}}.
\end{align}

In what follows, we will use the metric \eqref{10}, to study the tunneling of scalar particles from the event horizon. The tunneling rate is related to the imaginary part of the action in the classically forbidden region given by
\begin{equation}
\Gamma\sim\exp{\left(-\frac{2}{\hbar}\,\mbox{Im}\,I\right)}
\end{equation}

\section{Tunneling from SdS black hole with topological defects}
Let us now write the Klein-Gordon equation for a scalar field $\Psi $ given by
\begin{equation}
g^{\mu \nu }\partial _{\mu }\partial _{\nu }\Psi -\frac{m^{2}}{%
\hslash ^{2}}\Psi =0.  \label{15}
\end{equation}
where $m$ is the mass of the particle. We will use semi-classical WKB approximation, therefore we assume an ansatz of the form
\begin{equation}
\Psi (t,r,\theta ,\varphi)=e^{\left( \frac{i}{\hslash }I(t,r,\theta
,\varphi)+I_{1}(t,r,\theta ,\varphi)+O(\hbar )\right) },  \label{16}
\end{equation}

We can use the symmetries of the spacetime metric \eqref{10}, given by two Killing vectors, $\partial _{t}$ and $\partial _{\varphi}$. Now let us choose the following ansatz for the action 
\begin{equation}
I(t,r,\theta ,\varphi)=-(E_{b,p}-J_{b,p}\Omega)t+R(r)+H(\theta) +J_{b,p}\varphi, \label{17}
\end{equation}%
where $E_ {b, p}, $ $J_ {b, p} $ are the Komar's energy and angular momentum of the particle. Since a topological defects exists the energy and angular momentum are decreased by a factor of $p^ {2} b$. Therefore, the Komar's energy would be $E_{b,p}=(1-8\pi\eta^{2})(1-4\mu)E$ and similarly the angular momentum $J_{b,p}=(1-8\pi\eta^{2})(1-4\mu)J$ \cite{vilenkin,ren}. Evaluating Eqs. \eqref{15}, \eqref{16} and \eqref{17} in leading order of $\hbar$ and dividing by the exponential term and multiplying by $\hbar^{2}$ we get
\begin{equation}
0=-\frac{(\partial
_{t}I)^{2}}{A(r)}+B(r)(\partial
_{r}I)^{2}+\frac{(\partial _{\theta }I)^{2}}{C^{2}(r)}+\frac{(\partial _{\varphi}I)^{2}}{D^{2}(r)}+m^{2}. \label{18}
\end{equation}

Since we are interested in the radial trajectories, we can use the last equation and the values of the components $g^{\mu \nu}$, of the metric \eqref{10}, yielding
\begin{eqnarray}
R_{\pm}(r)=\pm \int \frac{\mathrm{d}r}{\sqrt{A(r)B(r)}}
\left[(E_{b,p}-J_{b,p}\Omega)^{2}-A(r)\left(\frac{(\partial_{\theta}H)^{2}}{C^{2}(r)}+\frac{J_{b,p}^{2}}{D^{2}(r)}+m^{2}\right)\right]^{\frac{1}{2}} \label{19}
\end{eqnarray}

Putting $\theta=\theta_{0}$ in the above equations, it follows that the emitted particles should not have motion in $\theta$-direction, therefore we will focus only on the radial trajectory. Integrating around the pole $r_ {H} $ we get
\begin{equation}
R_{\pm}(r_{H})=\pm\frac{\pi i r_{H}^{2}p^{2}b\,(E-J\Omega_{H})}{\Delta_{,r}(r_{H})}
\end{equation}
where $+/-$ correspond to the outgoing/ingoing solutions. But, at the event horizon the dragging angular velocity vanishes, i.e., $\Omega_{H} ({r=r_ {H}}) =0$, so the last equation simplifies to
\begin{equation}
R_{\pm}(r_{H})=\pm\frac{\pi i r_{H}^{2}p^{2}b\,E}{\Delta_{,r}(r_{H})}.\label{22}
\end{equation}

On the other hand, the probabilities of crossing the horizon for scalar particles in each direction are given by
\begin{equation}
P_{emission}\sim \exp \left( -\frac{2}{\hbar}\,\text{Im} I\right) =\exp \left( -\frac{2}{\hbar}\,\text{Im}R_{+}\right),
\end{equation}
\begin{equation}
P_{absorption}\sim \exp \left( -\frac{2}{\hbar}\,\text{Im}I\right) =\exp \left(
-\frac{2}{\hbar}\,\text{Im}R_{-}\right).
\end{equation}

Clearly, the probability of particles tunneling from inside to outside the horizon, by using  $R_{+}=-R_{-}$, reads
\begin{equation}\nonumber
\Gamma=\frac{P_{emission} }{P_{absorption} }=\frac{\exp \left( -2\,\text{Im}R_{+}\right)}{\exp \left( -2 \,\text{Im}R_{-}\right)}=\exp \left( -4\,\text{Im}R_{+}\right)
\end{equation}
where we have set $\hbar=1$ in the last equation. Taking the imaginary part of \eqref{22}, and using \eqref{6}, it follows
\begin{eqnarray}\nonumber
\mbox{Im}R_{+}& = &\frac{\pi r_{H}^{2}p^{2}b\,E}{\Delta_{,r}(r_{H})}\\
&\approx & 2\pi M p^{2}b\left(1+\frac{16M^{2}}{l^{2}}\right)E.
\end{eqnarray}

Then, the probability of particles tunneling from inside to outside will be
\begin{equation}
\Gamma=\exp\left[-8\pi M p^{2}b\left(1+\frac{16M^{2}}{l^{2}}\right)E\right].
\end{equation}

In order to find the Hawking temperature we have to compare the tunneling rate with the Boltzmann factor, $\Gamma=\exp{\left(\beta \,p^{2}b\,(E-\Omega_{H} J)\right)}$, where $\beta=1/T$. For the Hawking temperature it follows
\begin{equation}
T_{H}=\frac{1}{8\pi M}\left(1-\frac{16 M^{2}}{l^{2}}\right).
\end{equation}

From the last two equations one can see that actually the radiation deviate from pure thermal spectrum, as a consequence there is a correction to the Hawking temperature of SdS black hole. However, the temperature is independent of the presence of topological defects. In the particular case, when $l\to\infty $, i.e., $\Lambda=0$, the Hawking temperature reduces to Schwarzschild black hole temperature.

\section{Tunneling from RNdS black hole with topological defects }
The line element of the Reissner-Nordstr\"{o}m black hole with positive $\Lambda$ in the background spacetime with a spinning cosmic string and a global monopole is given by
\begin{eqnarray}\nonumber
\mathrm{d}s^{2}&=&-\left(1-\frac{2M}{r}+\frac{Q^{2}}{r^{2}}-\frac{r^{2}}{l^{2}}\right)\left(\mathrm{d}t+a\,\mathrm{d}\phi \right)^{2}+\left(1-\frac{2M}{r}+\frac{Q^{2}}{r^{2}}-\frac{r^{2}}{l^{2}}\right)^{-1}\mathrm{d}r^{2}\\
&+&r^{2}p^{2}\left(\mathrm{d}\theta^{2}+b^{2}\sin^{2}\theta\mathrm{d}\phi^{2}\right)\label{28}
\end{eqnarray}

Solving for $r^{4}-l^{2}r^{2}+2Ml^{2}r-l^{2}Q^{2}=0$, we can get the event horizon $r_{H}$ and the cosmological horizon $r_{C}$ location. Without getting into details (see, e.g., \cite{rahman1}), after expanding $r_{H}$ in terms of $M$, $Q$ and $l$, with $\Xi<1/27$, it was found 
\begin{equation}
r_{H}=\frac{1}{\alpha}\left(1+\frac{4M^{2}}{l^{2}\alpha^{2}}+\hdots\right)\left(M+\sqrt{M^{2}-Q^{2}\alpha}\right)\label{29}
\end{equation}
where $\alpha=\sqrt{1+4Q^{2}/l^{2}}$. By following the same arguments that have been used in the last section, we can define $\tilde{\Delta}=r^{2}+Q^ {2} -2Mr^ {2} -r^ {4} /l^ {2} $, in this way the metric \eqref{28} near the horizon takes a similar form as \eqref{10}, since
\begin{equation}
\Delta_{,r}(r_{H})=\frac{\mathrm{d}\tilde{\Delta}}{\mathrm{d}r}\bigg|_{r=r_{H}}=2\left(r_{H}-M-2\frac{r_{H}^{2}}{l^{2}}\right).
\end{equation}

In order to work out the tunneling rate of charge particles, let us now write the charged Klein-Gordon equation for the scalar field $\Psi$, given by 
\begin{eqnarray}\nonumber
\frac{1}{\sqrt{-g}}\left( \partial _{\mu }-\frac{iq}{\hbar }A_{\mu
}\right)\left( \sqrt{-g}g^{\mu \nu}(\partial _{\nu
}-\frac{iq}{\hbar }A_{\nu })\Psi \right)-\frac{m^{2}}{\hbar
^{2}}\Psi =0,
\end{eqnarray}
here $q$ is the charge of the particle and $A_{\mu}$ the electromagnetic four potential. We can use WKB approximation and assume an ansatz of the scalar field $\Psi$ and action $I(t,r,\theta ,\varphi)$, similar to \eqref{18} and \eqref{17}. Evaluating term by term in the highest order of $\hslash $ and dividing by the exponential term and multiplying by $\hslash ^{2},$ we get
\begin{equation}
0=-\frac{(\partial
_{t}I-qA_{t})^{2}}{A(r)}+B(r)(\partial
_{r}I)^{2}+\frac{(\partial _{\theta }I)^{2}}{C^{2}(r)}+\frac{(\partial _{\varphi}I)^{2}}{D^{2}(r)}+m^{2}. \label{9}
\end{equation}

Similarly, we will focus only on the radial trajectories. Integrating the last equation and using the values of the metric components $g^{\mu \nu} $, gives
\begin{eqnarray}
R_{\pm}\left(r\right)=\pm \int \frac{\mathrm{d}r}{\sqrt{A(r)B(r)}}
\left[((E_{b,p}-J_{b,p}\Omega)+qA_{t})^{2}-A\left(\frac{(\partial_{\theta}H)^{2}}{C^{2}}+\frac{J_{b,p}^{2}}{D^{2}}+m^{2}\right)\right]^{\frac{1}{2}} \label{19}
\end{eqnarray}

Integrating the last equation around the pole $r_ {H} $, and use the fact that also the Komar's charge of the particles decreases $q\to\left(1-8\pi\eta^{2}\right)(1-4\mu) q$ leads to
\begin{equation}
R_{\pm}=\pm\frac{\pi i r_{H}^{2}p^{2}b\left[\left(E-J\Omega_{H}\right)+qA_{t}\right]}{\Delta_{,r}(r_{H})}.
\end{equation}

It is not difficult to see that the dragged angular velocity vanishes at the horizon, i.e., $\Omega_{H}(r=r_{H})=0$, yielding
\begin{equation}
R_{\pm}(r_{H})=\pm\frac{\pi i r_{H}^{2}p^{2}b(E+qA_{t})}{\Delta_{,r}(r_{H})}.
\end{equation}

Taking the imaginary part of this equation and using \eqref{29} gives
\begin{eqnarray}
\mbox{Im}R_{+}& = &\frac{\pi r_{H}^{2}p^{2}\,b\left(E+qA_{t}\right)}{\Delta_{,r}(r_{H})}\\\nonumber
&\approx &\frac{\pi}{2\alpha}\frac{\left(M+\sqrt{M^{2}-Q^{2}\alpha}\right)^{2}p^{2}b\left(E+qA_{t}\right)}{M(1-\alpha)+\sqrt{M^{2}-Q^{2}\alpha}}
\end{eqnarray}

Thus, the probability of particles tunneling from inside to outside the event horizon using $A_{t}=Q/r_{H}$, is given by
\begin{eqnarray}
\Gamma=\exp \Bigg\{-\frac{2\pi}{\alpha}\frac{\left(M+\sqrt{M^{2}-Q^{2}\alpha}\right)^{2}p^{2}b\,E}{M\left(1-\alpha\right)+\sqrt{M^{2}-Q^{2}\alpha}}
\left[1-\frac{\alpha\, e \,Q}{M+\sqrt{M^{2}-Q^{2}\alpha}}\left(1-\frac{4M^{2}}{l^{2}\alpha^{2}}+\hdots\right)\right]\Bigg\}
\end{eqnarray}

The Hawking temperature of charged scalar particles can simply be found by comparing the tunneling rate with the Boltzmann factor $\Gamma=\exp{\left(\beta\,p^{2}b\,(E-\Omega_{H}J)\right)}$, where $\beta=1/T$. For the Hawking temperature if follows
\begin{eqnarray}
T_{H}=\frac{\alpha}{2\pi}\frac{M(1-\alpha)+\sqrt{M^{2}-Q^{2}\alpha}}{\left(M+\sqrt{M^{2}-Q^{2}\alpha}\right)^{2}}
\end{eqnarray}

As a result, we conclude that there are corrections to the Hawking radiation of RNdS black hole, but the Hawking temperature remains independent of topological defects. In the particular case, setting $l\to\infty $, i.e., $\Lambda=0$, and $\alpha=1$, we recover the Hawking temperature of charged scalar particles for RNdS black hole without topological defects. Clearly, for uncharged particles $q=0$, the last equation reduces to Perkih's result.

\section{Conclusion}

In summary, we have studied the emission of scalar particles from Schwarzschild-de Sitter and Reissner-Nordstr\"{o}m de-Sitter black holes with a spinning cosmic string and a global monopole. We have solved the Klein-Gordon equations for charged/uncharged scalar particles by using the Hamilton-Jacobi method and WKB approximation. In order to find the Hawking temperature we have compared the tunneling probability of outgoing particles with the Boltzmann factor by taking into account that Komar's energy, angular momentum, and the charge of the particles decreases by a factor of $p^{2}b$ in spacetime with topological defects. As a result, we have shown that the Hawking temperature of these black holes is independent of topological defects for both cases.

\section*{Acknowledgement}

The author would like to thank the editor and the anonymous reviewer for the very useful comments and suggestions which help us improve the quality of this paper.

\end{document}